\documentclass[usenatbib,useAMS,preprint,psfig2]{mn2e}
\usepackage{times}
\usepackage{amsmath}
\usepackage{amssymb,amsfonts}
\usepackage{graphicx}
\usepackage{epstopdf}

\def\ii{\'{\i}}

\def\ve{\varepsilon}
\def\F{{\rm F}}
\def\gapprox{\lower.4ex\hbox{$\;\buildrel >\over{\scriptstyle\sim}\;$}}
\def\lapprox{\lower.4ex\hbox{$\;\buildrel <\over{\scriptstyle\sim}\;$}}
\def\be{\begin{equation}}
\def\ee{\end{equation}}
\def\bea{\begin{eqnarray}}
\def\eea{\end{eqnarray}}

\begin{document}
\title{Pair emission from bare magnetized strange stars}
\author[Melrose, Fok \& Menezes]
{D.B. Melrose\thanks{E-mail: melrose@physics.usyd.edu.au}, 
R. Fok\thanks{present address: Department of Physics, University of Oregon,
1371 E 13th Avenue, Eugene, OR 97403, USA}
and D.P. Menezes\thanks{permanent address: Depto de F\ii sica - CFM - Universidade Federal de Santa
Catarina  Florian\'opolis - SC - CP. 476 - CEP 88.040 - 900 - Brazil}
\\
School of Physics, University of Sydney, NSW 2006,
Australia}

\date{}

\pagerange{\pageref{firstpage}--\pageref{lastpage}} \pubyear{2006}

\maketitle

\label{firstpage}

\begin{abstract}
The dominant emission from bare strange stars is thought to be
electron-positron pairs, produced through
spontaneous pair creation (SPC) in a surface layer of electrons tied
to the star by a superstrong electric field. 
The positrons escape freely, but the electrons are directed towards
the star and  quickly fill all available states, such that their degeneracy
suppresses further SPC. An electron must be reflected and 
gain energy in order to escape, along with the positron.
Each escaping electron leaves a hole that is immediately filled by 
another electron through SPC. 
We discuss the collisional processes that produce escaping electrons.
When the Landau quantization of the motion perpendicular
to the magnetic field is taken into account, electron-electron collisions 
can lead to an escaping electron only through a multi-stage
process involving higher Landau levels. Although the available
estimates of the collision rate are deficient in several ways, it appears
that the rate is too low for electron-electron collisions to be effective.
A simple kinetic model for electron-quark collisions leads to an
estimate of the rate of pair production that is analogous to
thermionic  emission, but the work function is poorly determined.
\end{abstract}


\section{Introduction}

It was proposed by \cite{W84} that deconfined, strange quark matter 
(SQM), consisting of the $u$-, $d$- and $s$-quarks, is an absolutely 
stable form of matter. A strange star (SS) is a compact object composed of SQM. 
In a model for a SS \citep{AFO86}, the quarks are bound
by the strong force, rather than the gravitational force that binds other stars. At 
the surface of the star, where the quark density drops abruptly to 
zero, the electrons extend into a layer of thickness $\Delta 
z\sim10^3\rm\,fm$ above the surface, where there is a superstrong electric 
field that ties the electrons to the star. It has been suggested
that such a bare SS might be covered by a crust \citep{AFO86} of 
ordinary nuclear matter (and neutralizing electrons), but such a 
crust could be blown away as the SS forms  \citep{U97}, or be destroyed
by thermal effects \citep{Ketal95}. Thus, if SSs 
form, one expects them to be bare, in the sense that the surface is 
this thin layer of electrons. The estimated plasma frequency 
in the surface layer, $\omega_p\approx19\rm\,MeV$,
precludes emission of photons below this frequency. 
Concerning the emission from SSs,
\cite{AFO86} commented that the luminosity may be very high,
but added the proviso that only if heat can be supplied rapidly
enough. On the contrary, if heat cannot be supplied rapidly
enough, the emission from a SS may be strongly suppressed.
A SS may be a  `silver sphere' rather than black
body (BB) \citep{AFO86}. The estimated electric field
in the electron layer is
$E\approx5\times10^{19}\rm\,V\,m^{-1}$ and this corresponds to
$E\sim30E_c$, where $E_c =1.3\times10^{18}\rm\,V\,m^{-1}$ is 
the Schwinger electric field \citep{S51}. According to quantum 
electrodynamics an electric field is intrinsically unstable to decay
due to spontaneous pair creation (SPC) 
\citep{S51}. Although this effect is negligible for $E\ll E_c$,
it becomes extremely efficient for superstrong electric 
fields with $E\gapprox E_c$. This suggests that SSs may be
highly luminous in emission of pairs, with the luminosity not
restricted by the Eddington limit \citep{U98}. It might be remarked
that at the expected temperature, $T\approx10^{11}\rm\,K$, just
after a SS forms, BB emission is dominated by pairs,
with the power radiated in pairs being 7/4 times
the BB power radiated in photons in vacuo \citep{LL59}.
The luminosity due to SPC is not BB and once the star
reaches thermal equilibrium, the luminosity must
be suppressed to no more than the BB level.
More recent discussions of the emission spectrum have 
included thermal effects, emphasizing
interactions between escaping pairs and
photons \citep{U01,AMU03,AMU04}. 

The original motivation for the investigation reported here was
to generalize existing treatments of pair emission from 
a bare SS \citep{U98} to include the effects of a magnetic field. 
However, questions concerning the underlying
physics arose, and these are emphasized in the present paper,
specifically, self-regulation of SPC, the role of collisions,
and the sequence of processes that allows
an electron to escape.  \cite{U98} noted that unsuppressed SPC
would produce pairs at the rate
$\approx4\times10^{56}\rm\,s^{-1}$, but that such emission
could persist only for a short time. SPC creates
electrons propagating towards the star and positrons propagating
away from the star, and a steady state escape is possible
only if upward propagating electrons are available to accompany
the escaping positrons. The copious
source of downward propagating electrons from SPC fills any unoccupied
electron states, suppressing
SPC through the Pauli exclusion principle. 
No electron can be created by SPC if the state in which it would be
created is already occupied. Hence, once all relevant states
are filled, SPC is suppressed. In the absence of collisions,
SPC completely suppresses itself, so that the SS would be an ideal
silver sphere with no pair emission.  

Collisions play a central role in allowing electrons 
(and hence pairs) to escape. For escape to occur, a sequence of collisions
must transfer an electron from the region of phase space
where it was created by SPC into the region of phase space
corresponding to escape, that is, such that it has an upward directed
 motion at greater than the escape energy. The electron escapes, leaving a
hole in the distribution that is immediately filled by SPC, with
the newly created positron escaping with this electron.
The rate of pair emission is determined by the rate collisions
transfer electrons between these two regions of phase space.
In existing treatments, this rate is (implicitly) assumed arbitrarily high, so
that there is sufficient time for thermalization, allowing the electron
distribution to be approximated by a Fermi-Dirac (FD) distribution.

An important qualitative point is that the dominant role
of SPC implies that the electrons in the electron layer
cannot be in thermal equilibrium.
The difference, $\delta n$ say, between the electron
occupation number and unity regulates SPC, such that
SPC is reduced from its value in vacuo by the factor $\delta n$.
This suppression was discussed by \cite{U98} assuming a
FD distribution for the electrons. However, a FD
distribution applies only in thermal equilibrium, and the
presence of SPC invalidates this assumption. Granted
that the suppression of SPC must be self-regulatory, 
the value of $\delta n$ must adjust to allow pair creation at the
rate required to replace pairs that escape. The time scale for escape
is much shorter than the time scale for thermal equilibrium
to be reached in the electron layer, and the steady
state distribution is determined by collisions, SPC and escape,
and not by thermal equilibrium. In principle, escape can
result from a sequence of electron-electron collisions in the
electron layer, or from electron-quark collisions below this layer.
Both are discussed in this paper.

The presence of a magnetic field affects pair  emission from a 
bare SS in several ways. Typical surface magnetic fields in ordinary 
pulsars are of order $0.1B_c$,  where 
$B_c=m^2c^2/e\hbar=4.4\times10^9\rm\,T$ is 
the so-called critical magnetic field. The surface field of a SS 
is thought to be considerably higher \citep{AMU04}, of order 
$10B_c$, as in magnetars \citep{TD95}. 
Such a magnetic field affects the rate of SPC in vacuo.
However, granted that SPC
is regulated through the Pauli exculsion 
principle, this effect is relatively unimportant. A magnetic 
field also affects the electron energy states: free motion is 
restricted to one dimension (1D) along the magnetic field lines, with the 
perpendicular motion replaced by a set of discrete Landau levels. 
An electron in the $n$th Landau level acts like a 1D
particle with an  effective mass, $m_n$, that depends on the 
Landau level. This poses a problem if all electrons are in
their lowest Landau level: collisions in 1D between 
particles with the same rest mass cannot change the energies
of the particles, and hence electron-electron collisions at fixed $n$,
specifically both initial and both final particles with the same $n$,
cannot transfer electrons from one region of momentum
space to another, and so cannot contribute to the escape of pairs.
In order for an electron to escape as a result 
of electron-electron collisions, collision-induced jumps in Landau 
level are essential. There is no corresponding constraint on
electron-quark collisions.

In section~2 we discuss how SPC and the escape of electrons
are regulated by collisions. In section~3 we identify
the sequence of electron-electron collisions that allows escape.
In section~4 we consider the effects of electron-quark collisions. 
The results are summarized and discussed in section~5.

\section{Regulation of SPC through degeneracy}

The nonthermal nature of pair emission by a SS raises some
unfamiliar problems that are discussed from a qualitative
viewpoint in this section. For the purpose of discussion, the
magnetic field is assumed to be vertical, and the perpendicular
motion of the electrons is assumed to be quantized into the
Landau levels, $n=0,1,2,\ldots$. SPC is assumed to generate only downward
propagating electrons with $\ve<eV$ in the ground level, $n=0$.

\subsection{Suppression of SPC}

The decay of an electric field $E\gg E_c$ in vacuo
produces pairs at the Schwinger rate $a_{\rm SPC}(E/E_c)^2$
with $a_{\rm SPC}=1.7\times10^{56}\rm\,m^{-3}\,s^{-1}$
\citep{S51}. The generalization to an electric field along a magnetic field was given by
\cite{DL76}, and is written down in the Appendix~A. The pair creation
produces electrons only in a restricted region of momentum
space, centered on momenta anti-parallel to the electric field, with 
energy $\ve\le eV$. It is convenient to introduce the density of states factor 
$D_{\rm SPC}=\int d^3{\bf p}/(2\pi\hbar)^3$, where the integral is over the region
of momentum space to which SPC is
restricted. The quantity $D_{\rm SPC}$ is estimated below, cf.\ (\ref{DSPC}), for
the magnetized case. The pair creation rate in a degenerate electron
gas is suppressed by a factor $1-n({\bf p})$, where $n({\bf p})$ is the
occupation number of the created electron. In the absence of
such suppression, SPC would
lead to a SS luminosity in pairs of $\approx3\times10^{44}\rm\,W$;
this may occur just after the formation of a SS when its surface
is very hot, but could 
persist only for a very brief period \citep{U98}. In treating this suppression,
 \cite{U98} assumed that 
$n({\bf p})=n_{\rm FD}({\bf p})$ has its thermal (FD) value. 
The value of $n({\bf p})$ in the region
of momentum space where the electrons are created results from a
balance between gains and losses, and a FD distribution results only
if both gains and losses are dominated by collisions. Creation of
electrons implies $n({\bf p})>n_{\rm FD}({\bf p})$ in this momentum
range, and the suppression factor, $\delta n({\bf p})=1-n({\bf p})$, is then
smaller than the thermal value, $1-n_{\rm FD}({\bf p})$. 

The kinetic equation for electrons in the region of momentum 
space where SPC operates is
\be
\frac{dn({\bf p})}{dt}=\nu_{\rm SPC}({\bf p})\delta n({\bf p})
-\nu_{\rm loss}({\bf p})\,n({\bf p}),
\label{dndt}
\ee
with $\nu_{\rm SPC}({\bf p})=a_{\rm SPC}(E/E_c)^2/D_{\rm SPC}$
and where $\nu_{\rm loss}({\bf p})$ 
includes collisional losses and escape of electrons. 
When the suppression is strong, one has
$n({\bf p})\approx1$, 
$\delta n({\bf p})\approx\nu_{\rm loss}({\bf p})/\nu_{\rm SPC}({\bf p})$.
SPC is then adjusted to the value required to
balance the escape of pairs from the star. 

\subsection{Role of collisions}

When SPC is the dominant effect on the distribution of electrons in the
electron layer, the distribution function may be determined by first
neglecting collisions and thermal motions, to find a zeroth order
distribution, and then including collisions and thermal motions as
perturbations.

In the absence of collisions and thermal motions, SPC in the electron layer
fills all the electron states corresponding to downward motion with $\ve<eV$, $n=0$. 
Each electron propagates downward until it encounters a quark and is reflected. 
In the absence of thermal motions, the quark is stationary, and due to the
large mass ratio, the electron is reflected with no change in energy,
resulting in all the upward electrons states with $\ve<eV$, $n=0$
also being filled. Electrons oscillate across the electron layer, with downgoing
electrons reflected off quarks below the electron layer and with upward electrons
reflected at the top of the layer by the electric field. Thus, the zeroth order
distribution of electrons in the electron layer corresponds 
to occupation number equal to unity for $\ve<eV$ and zero for $\ve>eV$,
with no electrons in Landau levels with $n>0$.

Now consider the effect of collisions, which
tend to thermalize the distribution. A thermal distribution is a FD
distribution, $n(p)=1/(\exp\{[\varepsilon-\mu_e]/T\}+1)$,  
where $\mu_e$ is the  chemical potential, and $T$ is the 
temperature in energy units. Provided the Fermi
energy satisfies $\ve_\F\gg T$, one has $\mu_e\approx\ve_\F$. 
A FD distribution has a Boltzmann-like tail such that there is a probability
$\propto\exp(-\ve/T)$ of finding an electron with $\ve\gg\ve_\F$. Because upgoing 
electrons with $\ve>\ve_{\rm esc}$ escape directly, even if collisions
produce a thermal distribution at lower energies, this escape leaves a depleted
high-energy tail. Collisions tend to restore a thermal distribution by
feeding electrons into the depleted tail. The rate at which escaping
electrons are produced is determined by the rate collisions feed
electrons into the tail \citep{G60}. The relevant collisions can be either
electron-electron collisions in the electron layer, or electron-quark
collisions below this layer. 

This discussion presupposes that there is a magnetic field present
so that the perpendicular energy is quantized. In the absence of a magnetic
field, SPC produces electrons only parallel to $\bf E$. Collisions tend to
scatter the electrons into other directions, so that they diffuse in angle,
leading to an increase in the angular spread. Discussion of this effect
is more complicated that discussion in the case where the perpendicular
energy is quantized case, and  we restrict our discussion to the quantized case. 

\subsection{Transitions between Landau levels}

The zeroth order distribution function, generated by SPC, is
one dimensional (1D), along ${\bf B}$. Electron-electron collisions in 1D do not
change the energies of the electrons. Collisions between electrons
in different energy levels, that remain in those levels, can 
redistribute parallel momentum. This may be understood by
noting that the electrons in different Landau levels act like particles
with different rest masses.

Let $p$ be the component of the momentum 
along the magnetic field. The energy of an electron in
the $n$th Landau level is
\begin{equation}
\varepsilon_n(p) = (m_n^2c^4 + p^2c^2)^{1/2},
\qquad
m_n=m(1+2nB/B_c)^{1/2}.
\label{electronenergy}
\end{equation}
The form (\ref{electronenergy}) implies that 
electrons act like 1D particles with a rest mass, $m_n$.
The presence of an electric field along the magnetic field does not
affect the Landau levels.

\subsection{Electron distribution in a magnetized electron plasma}

The zeroth order distribution function, identified above, 
has occupation number equal to unity in the ground state
for $\ve<eV$, with no electrons with $\ve>eV$, and none 
in higher Landau levels. This corresponds formally
to a FD distribution in the limit $T\to0$, $B\to\infty$. Thermal contact
between the electron layer and the quark layer below it
tends to drive this zeroth order distribution towards a
FD distribution at the temperature of the quarks. Thus collisions
tend to transfer electrons from $\ve<\ve_\F$, where the
zeroth order distribution is overpopulated compared with
a FD distribution, to $\ve>\ve_\F$, where it is
underpopulated. Similarly, any tendency to
thermalization causes the occupation number
of the higher Landau levels, $n\ge1$, to be nonzero.

A FD distribution, $n_n(p)$,
for electrons in the $n$th Landau level is
$n_n(p)=1/(\exp\{[\varepsilon_n(p)-\mu_e]/T\}+1)$. 
The Fermi energy, $\ve_\F$, is the same for all Landau levels, but 
the Fermi momentum, $p_{\F n}$, depends on $n$, with $n_n(p)$ 
non-negligible only for $-p_{\F n}\lapprox p\lapprox p_{\F n}$. One 
has
\be
p_{\F n}=(\ve_\F^2/c^2-m_n^2c^2)^{1/2}.
\label{pFn}
\ee
There is a maximum Landau level, $n=n_{\rm max}$, such that $p_{\F n}$, as 
defined by (\ref{pFn}), is real for $n<n_{\rm max}$ and imaginary for 
$n>n_{\rm max}$. Assuming $\ve_\F=eV\gg mc^2$, one finds
\be
2n_{\rm max}\frac{B}{B_c}\approx\left(\frac{eV}{mc^2}\right)^2.
\label{nmax}
\ee
The number density of electrons, $N_e$, assuming that all Landau levels up to
$n_{\rm max}$ are filled is then
\be
N_e=\sum_{n=0}^{n_{\rm max}}4\frac{eB}{2\pi\hbar}\frac{p_{\F n}}{2\pi\hbar}
\approx8\pi\left(\frac{eV}{2\pi\hbar c}\right)^3.
\label{Ne}
\ee
The final expression in (\ref{Ne}) is the same as for a degenerate
FD distribution in the absence of a magnetic field. 

\section{Electron-electron collisions}

In this section we assume that the 
production of escaping pairs is determined by
electron-electron collisions in the electron layer. 
However, there are sufficient uncertainties that
it is unclear whether or not the rate
of electron-electron collisions is high enough for
this to be the case.  

\subsection{Kinematics of electron collisions in a magnetic field}

Conservation of energy and momentum in a collision places a severe 
restriction on collisions that can lead to an electron escaping. Let 
the initial and final values of $p$ be $p_i$, with $i=1,2$ for the 
initial electrons, and $i=3,4$ for the final electrons. The $i$th 
electron is assumed to be in the Landau level $n_i$. Electrons in 
different Landau levels effectively have different rest masses, with 
the effective rest mass for the Landau level $n_i$ being $m_{n_i}$, 
cf.\ (\ref{electronenergy}). To simplify the notation we write the effective 
mass of the $i$th particle as $M_i$, with $M_i=m_{n_i}$. 
Qualitatively, the kinematics are equivalent to those for collisions 
between (relativistic) particles of different mass constrained to 
move in 1D. Conservation of momentum implies $P=p_1 + p_2 = 
p_3 + p_4$, and conservation of energy implies 
$E=\varepsilon_{n_1}(p_1) + \varepsilon_{n_2}(p_2) = 
\varepsilon_{n_3}(p_3) + \varepsilon_{n_4}(p_4)$. There are two
solutions for the final variables, and we solve for
$p_3$ and $\ve_3$, with $p_4=P-p_3$ and $\ve_4=E-\ve_3$. 
The solutions are
\begin{equation}
p_3 = \frac{D_1\,P\pm D_2E/c}{2M^2},
\quad
\ve_3 = \frac{D_1\,E\pm D_2Pc}{2M^2},
\label{p3}
\end{equation}
with $M^2=(E/c^2)^2-(P/c)^2$ and with
\begin{equation}
D_1= M^2 + M_3^2 - M_4^2,
\quad
D_2=[D_1^2-4M^2M_3^2]^{1/2}.
\label{D12}
\end{equation}

\subsection{Collision leading to escape}

For an electron to escape as the result of a sequence of collisions,
its energy must increase from below the escape threshold to above it. 
The first step in the sequence must be a collision in 
which one or both initial electrons are in the lowest Landau level, with 
both final electrons in higher Landau levels, $n>0$. 
The electron-electron collision rate \citep{L81,SM87} is a rapidly decreasing 
function of the differences between the Landau levels of the initial 
and final electrons. This
implies that the changes in Landau level occur preferentially in 
single steps, with say $|n_3-n_1|=1$ and $n_4=n_2$.
Using (\ref{p3}) and (\ref{D12}), one finds that
collisions in which the Landau level increases, specifically $n_3=n_1+1$
in the present case, lead to a decrease in the kinetic energy, which
may be attributed to a conversion of kinetic energy into rest energy.
Such scattering impedes rather than facilitates
electrons escaping. On the other hand,
scattering from higher to lower Landau levels, $n_3=n_1-1$
in the present case, effectively converts rest energy 
into kinetic energy and can result in an electron with sufficient energy escaping.  

A sequence of collisions that leads to escape is as follows. First, an initial
electron with $n_1=0$, $-eV/c<p_1<eV/c$ is scattered to $n_3=1$ by
colliding with an electron with $n_2=n_4\ge1$. Second, collisions between
electrons with $n_1,n_2\ne0$, $n_1\ne n_2$ allow the energies of the electrons
to change, feeding electrons into the higher energy states. Third, the final collision
is between an electron
with $n_1=1$, $\ve_1\approx eV$ and an electron with 
$n_2>1$, $\ve_2\approx eV$ that results in an electron with
$n_3=0$, $\ve_3>eV$ that can escape.

In a steady state the SPC rate must be equal to 
the escape rate of pairs. The transfer rate of electrons from the 
state in which they are created to the final state from which they 
escape is regulated by the value of $\delta n$ in the 
intermediate states. The following 
semi-quantitative argument indicates how this regulation occurs.

\subsection{Steady state solution for higher Landau levels}

Consider a Landau level $n_1$. Collisions transfer electrons into 
this level from other levels, and out of this level to other levels. 
Ignoring SPC and escape, the
net rate at which electrons enter the $n_1$th level with momentum 
$p_1$ is
\begin{equation}
\frac{dn_{n_1}(p_1)}{dt} = \int \frac{dp_2}{A} \sum_{n_2,n_3,n_4} [ 
G_{34}^{12} - G_{12}^{34}],
\label{rate1}
\end{equation}
with
\begin{eqnarray}
G_{34}^{12} - G_{12}^{34} = r_{34}^{12} \{ 
n_{n_3}(p_3)n_{n_4}(p_4)\delta n_{n_1}(p_1) \delta n_{n_2}(p_2)
  \nonumber \\
- n_{n_1}(p_1)n_{n_2}(p_2)\delta n_{n_3}(p_3)\delta n_{n_4}(p_4)\},
\label{rate2}
\end{eqnarray}
with $\delta n_k = 1-n_k$, and where $A$ is a normalization constant; 
$r_{34}^{12}$ is the collision rate between electrons in initial 
and final Landau levels $n_1,n_2$ and $n_3,n_4$. The collision rate 
is symmetric under the interchanges $1\leftrightarrow2$, $3\leftrightarrow4$ 
and $12\leftrightarrow34$. In a 
steady state one has $dn_{n_1}(p_1)/dt=0$ for all $n_1$ and $p_1$. 
The kinetic equation (\ref{rate1}) needs to be modified for the 
ground state, $n=0$, to take into account SPC and 
escape of electrons. For the present we ignore the ground state.

We consider only transitions that involve one electron changing
Landau level by unity. For simplicity, we assume that all transitions occur at the 
same rate, $R$ say, and that all the occupation numbers are nearly 
equal to unity, $\delta n_{n_i}=1-n_{n_i}\ll1$ for energies below the 
Fermi energy. Then, for $n_1>0$, (\ref{rate1}) is replaced by
\begin{equation}
\frac{dn_{n_1}}{dt}=R[\delta n_{n_1}\, \delta 
n_{n_2}-\delta n_{n_3}\, \delta n_{n_4}],
\label{rate3}
\end{equation}
where the arguments $p_i$ are redundant in this simplified model. A 
steady state requires $dn_{n_1}/dt=0$, and in this simplified 
model this condition is satisfied if all the occupation numbers are 
equal, $\delta n_{n_i}=\delta n_1$, below the Fermi energy. 

\subsection{Collisions involving the ground state}

Now consider the kinetic equation for the ground state. 
In the region, $-eV/c<p<0$, where SPC occurs, $\delta n_0=1-n_0\ll1$ is
determined by balancing the rate of increase due to 
SPC with the rate of decrease due to collisions. 
According to the foregoing model, collisions that
transfer electrons from $n_1=0$ to $n_3=1$ cause
losses to the ground state at
the rate $-R(\delta n_1)^2$. 
The rate of increase in $n_0$ due to SPC is 
$\nu_{\rm SPC}\delta n_0$ where
$\nu_{\rm SPC}$ is the intrinsic rate of SPC, given by
\be
\nu_{\rm SPC}=\frac{w_\parallel}{D_{\rm SPC}},
\label{nuSPC}
\ee
with $w_\parallel$ given by (\ref{Edecay}), and the density of states factor by
\be
D_{\rm SPC}=\frac{eB}{2\pi\hbar}
\int_{-eV/c}^0\frac{dp}{2\pi\hbar}
=2\pi\frac{B}{B_c}\,\frac{eV}{mc^2}\left(\frac{mc}{2\pi\hbar}\right)^3.
\label{DSPC}
\ee
The model is valid only for $\delta n_0\ll\delta n_1$, and then one has
$\delta n_0=R(\delta n_1)^2/\nu_{\rm SPC}\ll1$. In the opposite limit, 
collisional effects dominate and lead to a FD distribution.

The collisions that lead to escape involve an electron with $n=1$,
$\ve\lapprox\ve_\F$ being scattered to $n=0$, $\ve>\ve_\F$. 
The most favorable case is for $p_1=p_2=p_{\F1}$, when one
has $P=2p_{\F1}$, $E=2\ve_\F$.
According to (\ref{p3}) and (\ref{D12}), with $n_1=n_2=n_4=1$, 
$n_3=0$, the $+$-solutions has $p_3>p_\F$, $\ve_3>\ve_\F$,
implying that the electron can escape. There is a small range of
$p_1$ near $p_{\F1}$ that allows $p_3>p_\F$, and this implies that
there is a fraction of collisions that result
in an escaping electron, with this fraction, $\Delta_e$ say,
being of order the ratio of this small momentum range to $p_{\F1}$.
The rate escaping electrons are produced is then
$\Delta_eR\delta n_1$. This rate must be equal to the
rate, $\nu_{\rm SPC}\delta n_0$, of SPC. Equating the rates then
gives $\delta n_1=\Delta_e$, $\delta n_0=R(\Delta_e)^2/\nu_{\rm SPC}$.

This simple model establishes in principle how the system
can regulate SPC to balance the rate of escape of pairs.
The important qualitative point is that the pair creation
rate and the escape rate are determined
purely by collisional effects. These rates are equal 
to $R(\Delta_e)^2$ in this simple model.

\subsection{Scattering rate}

Collisions in the electron layer determine the net rate of 
SPC only if a typical electron has many collision
before reaching the lower surface of the electron layer.
The collision rate was estimated by \cite{U01} using the
results of \citep{Petal99}:
\be
\nu_{ee}\approx A\frac{\ve_\F}{\hbar}h(\zeta),
\quad
A\approx1.3\times10^{-5},
\label{nuee}
\ee
with $\zeta\approx0.1\ve_\F/T$, and
$h(\zeta)\approx51/\zeta^2$ for $\zeta\gg20$ and
$h(\zeta)\approx(\zeta/3)\ln(2/\zeta)$ for $\zeta\lapprox1$,
with interpolation formulas at intermediate values. The
maximum value of $\nu_{ee}$ is of order 
$A\ve_\F/\hbar\approx3\times10^{17}\rm\,s^{-1}$. 
An electron propagates through the electron layer
in a time $\approx3\times10^{-21}\rm\,s$, suggesting that a
typical electron experiences a probability of a collision of order $10^{-3}$
in one traversal of the electron layer.  We conclude that electron-electron
collisions in the electron layer can govern the rate that
escaping pairs are produced only if (\ref{nuee}) seriously
underestimates the collision rate. 

The rate (\ref{nuee}) neglects several important effects, and needs to be modified to
take them into account. We comment on three separate effects that might
lead to (\ref{nuee}) being a serious underestimate of the actual rate. The
three effects are:
electromagnetic rather than electrostatic interactions for
relativistic particles, the nonthermal form of the distribution
function, and the quantization of the perpendicular energy for
a magnetized electron.

The collision rate (\ref{nuee}) is calculated assuming that the
interaction between the two electrons is longitudinal, with the
interaction cut off at distances greater than the Debye length. 
However, for highly relativistic particles, the dominant interaction
is transverse rather than longitudinal, and this is not affected by
Debye screening \citep{HP93}. As a result
the effective electron-electron collision rate is higher than
estimated by \cite{U01}. \cite{HP93} estimated that the contribution
from the transverse response exceeds that from the longitudinal response
by a factor of $0.70(\hbar cq_D/kT)^{1/3}$
for $kT\ll \hbar cq_D$, where $q_D$ is the Debye wave number. With
$q_D=(\omega_p/c)(kT/mc^2)^{-1/2}$, this factor becomes   
$0.70(\hbar cq_D/kT)^{1/3}\approx1.7(\hbar\omega_p/20\rm\,MeV)^{1/3}(kT/{1\rm\,MeV})^{-1/2}$
for $kT\ll\hbar\omega_p$. This factor is large only for $kT\ll1\rm\,MeV$, whereas
the relativistic assumption requires $kT\gg1\rm\,MeV$. Hence, although
the neglect of the transverse response made lead to (\ref{nuee}) being
an underestimate, the underestimattion is probably not by a very large factor.

A second effect that needs to be taken into account in revising (\ref{nuee})
for the present application is that the occupation numbers are
determined primarily by the collisions themselves, and not by thermal effects.
In a degenerate thermal electron gas, the combination of occupation
numbers in (\ref{rate2}) effectively restricts the momentum integral
to a small fraction $\sim kT/\ve_\F$ where the occupation number is
not close to either unity or zero, so that the collision rate is
suppressed except for electrons around the top of the Fermi sea.
When the occupation numbers are determined
by the collisions themselves, they approach
thermal values only if the transfer between 
states is dominated by collisions, and this is not the case when the
dominant effect is a flow from a source (SPC) in one region of
momentum space to a sink (escape) in another. This
should also cause (\ref{nuee}) to be an
underestimate of the actual collision rate. However, to estimate
the magnitude of this effect requires a detailed analysis that is
not attempted here.

The magnetic field affects the scattering in that the perpendicular energy must
change by discrete amounts, associated with transitions between Landau levels.
In (\ref{nuee}), the perpendicular energy levels are assumed continuous, and an
integration is performed over relevant angular variables.
Although the relativistic quantum theory of electron-electron scattering
in a magnetized plasma is available  \citep{L81,SM87}, only a few special 
cases have been calculated \citep{L81} in detail. 
With $B/B_c$ of order unity or greater, the transition rate 
between Landau levels decreases rapidly with the net change 
in the Landau quantum numbers, and changes by more than 
unity become increasingly unfavorable with increasing $B/B_c$. 

We conclude that although existing estimates of the collision rate 
suggest that it is too low for collisions within the electron layer to
affect the distribution function substantially, these estimates are
for the unmagnetized thermal case, and the thermal assumption
in particular can lead to a serious underestimate of the effective
rate. A detailed analysis is  required to determine whether
collisions in the electron layer alone can determine the rate of
production of escaping pairs.

\section{Electron-quark collisions}

In this section we ignore collisions in the electron
layer and consider the effect of collisions between
quarks and electrons below the electron layer.

\subsection{Reflected nonthermal electrons}

The downward propagating electrons, with a
nonthermal distribution due to SPC, are scattered by
quarks and electrons below the surface. As in the
electron layer, electron-electron collisions that do not
change the Landau level do not change
the occupation number. However, a collision
with a quark can change the electron energy
without changing its Landau level. We concentrate on
such collisions, which reflect electrons (along the magnetic field) with small
changes in energy. The fractional change in electron energy is of order the ratio of the 
thermal speed of the quark to the speed of light. 
A collision is only possible if there is an unoccupied
state available to the reflected electron.
This restricts effective collisions to near the maximum
energy, $eV$, of the electrons due to SPC. The distribution
function is modified from a step function at $\ve=eV$ through the
formation of a high energy tail. Collisions cause electrons to diffuse
in energy from $\ve<eV$ to $\ve>eV$ to populate this tail. This
results in escaping electrons when the tail extends to the
escape energy, $\ve_{\rm esc}$.

\subsection{Energy flux into the tail}

The electron distribution formed by SPC alone has the form of
a completely degenerate distribution with Fermi energy $eV$.
In the presence of thermal quarks, at a temperature $T$, this
distribution may be regarded as having a depleted (actually absent)
high-energy tail. The argument given by \cite{G60} implies that
collisions tend to feed electrons into the tail to establish a thermal
distribution, in this case a FD distribution at temperature $T$
(and chemical potential $eV$). In the case considered by
\cite{G60} the tail is depleted by an acceleration mechanism
transferring the particles to higher energy, whereas in the present
case the depletion is due to electrons with $\ve>\ve_{\rm esc}$
escaping. A modification of the method to treat this case is outlined
in Appendix~B. It implies that there is a flux in energy space that
corresponds to a flow from an unspecified source at low energy
to the sink at $\ve=\ve_{\rm esc}$. The collisions
described by a collision frequency, $\nu_c$, and a momentum-dependent
diffusion coefficient, $d(p)$, with $p=\ve/c$ in the ultrarelativistic case
assumed here. The analysis in Appendix~B implies that the rate
per unit volume and per unit time that electrons are fed into the range
$\ve=\ve_{\rm esc}$ where they escape is $\nu_cN_e$ times a
dimensionless factor that involves an integral over $1/d(p)$.
If the collisions are treated as 1D reflections off hard spheres,
this factor becomes $\exp(-W/kT)$, with
\be
W=\ve_{\rm esc}-eV
\label{wf}
\ee
the energy gap between the top of the Fermi sea and the escape
energy. The model has no spatial dependence, and is
concerned only with the flow of electrons in energy space
to restore a depleted high-energy tail.

The assumption that the collisions of electrons with quarks may be treated as 1D reflections 
off hard spheres leads to a Fermi-like acceleration of the electrons.
The electrons diffuse in $p$, and tend to gain energy because
head-on, energy-increasing collisions are slightly more frequent
than overtaking, energy-decreasing collisions. The diffusion
coefficient is independent of $p$, which leads to (\ref{wf}).

Collisions cause both diffusion in energy space and diffusion
in coordinate space. In a collision time an electron propagates
a mean free path, $c/\nu_c$, and in a typical collision, the energy of
an electron changes by $\delta\ve\sim\ve_\F(kT/m_qc^2)^{1/2}$, 
with $\ve_\F=eV$, and where $m_q$ is the mass of a quark.
An electron diffuses $W/\delta\ve$ mean free paths in the time
it take to diffuse through a range $W$ in energy space. This
provides an estimate of the thickness of the layer from which
electrons escape: $(c/\nu_c)/(W/\delta\ve)$. The rate per unit
time and per unit area that electrons escape from the surface
of the star is given by multiplying $\nu_cN_e\exp(-W/kT)$ by
this thickness. The result is independent of the collision
frequency: the thickness of the layer increases with decreasing
$\nu_c$ and compensates for the reduced rate electrons are
fed into the depleted tail.

\subsection{Thermionic-like emission}

The escape of electrons from the surface of a SS due to
collisions with thermal quarks is somewhat analogous to 
thermionic emission from a metal surface, sometimes called the
Richardson effect. Analogous features are that both involve
work functions, $W$, with $W$ given by (\ref{wf}) in the present
case, and both correspond to escaping electrons in a high-energy tail, 
where a FD distribution may be approximated by a Boltzmann
distribution. Also in both cases, collisions are required to
replenish the depleted high-energy tail, but the rate of
escape does not depend explicitly on the collision frequency.
A notable difference is that, in the case of SSs, the
electrons are highly relativistic. 

The estimated rate at which electrons escape from the 
surface of the SS is
\be
R_{\rm Th}=
4\pi R_*^2cN_e\,\frac{W}{eV}
\left(\frac{kT}{m_qc^2}\right)^{-1/2}
\,e^{-W/kT},
\label{Richardson}
\ee
where $R_*$ is the radius of the star, and with $W$ given by (\ref{wf}).
However, the actual value of $W$ is not well-determined due to
uncertainties in the estimate of the difference between the electric
potential energy $eV$ across the electron layer and the escape energy
$\ve_{\rm esc}$.

An upper bound on the power radiated in pairs is the BB limit.
BB emission at $T\gg10^9\rm\,K$ includes a luminosity
in pairs that is 7/4 times the photon luminosity in vacuo.
Thus the thermal rate is $7\pi R_*^2\sigma_{\rm SB}T^4$,
where $\sigma_{\rm SB}$ is the Stefan-Boltzmann constant.
The estimate (\ref{Richardson}) applies only if $R_{\rm Th}$
is much less than this thermal rate, in which case the star
acts like a silver sphere rather than a BB.

\section{Discussion and conclusions}

Our initial motivation for this investigation was to generalize the model of
\cite{U98} for the emission of pairs from SSs to include the effect of a
magnetic field. The most important qualitative effect of the magnetic field
is that it leads to quantization of the electron motion perpendicular to
the magnetic field into discrete Landau levels, and SPC creates electrons
only in the lowest Landau level. However, the investigation raised
questions relating to the role of collisions and of thermalization that are
the main topics discussed in this paper.

An important qualitative point is whether the electrons distribution in the
surface layers of a SS can be treated as a thermal distribution. There are
two processes that drive the electrons away from a thermal distribution:
SPC and escape of electrons. SPC creates electrons in one region of
phase space, and electrons can escape only from another region of
phase space. Collisions must drive the flux (in energy space) of electrons
between this source and sink. The electron distribution can be assumed
approximately thermal only if these effects can be treated as perturbations.
We argue that the reverse is the case: the dominant effects are SPC and
escape, and the tendency of collisions to thermalize the distribution 
can be treated as a perturbation. In the absence of collisions, the
electron states available to SPC are all filled, SPC is completely
suppressed and the SS is an ideal silver sphere. We discuss two types
of collision process that allows SPC to occur and electrons to escape:
electron-electron collisions and electron-quark collisions.

Electron-electron collisions can lead to an escaping electron through a
three-stage process. The first stage involves excitation of an electron
from the lowest Landau level at $\ve<eV$, where SPC can operate,
to a higher Landau level. The vacated state is immediately filled by
SPC, with the positron escaping. 
The second stage involves collisions between electrons in different
Landau states $n\ge1$ that redistribute the parallel
energy amongst the electrons in each of these levels.
The third stage involves a collision that transfers the electron back to
the ground state, $n=0$, above the escape energy. 
Scattering to higher Landau levels effectively
converts kinetic energy into rest energy, and scattering from
higher to lower Landau levels effectively converts rest energy
into kinetic energy. Only the latter allows the kinetic energy to
increase, as required to produce an escaping electron.
This sequence of electron-electron collisions competes with collisions
between quarks and electrons below the electron layer in producing
escaping electrons, and it can be the dominant effect only if the
electron-electron collision rate is high enough. With the
collision rate given by \cite{U01}, this is not the case, with a typical electron traversing the
electron layer many times before experiencing an electron-electron collision.
However, the estimate used by \cite{U01} needs revision
due to at least three effects.
First, it assumes a Debye screened longitudinal interaction, whereas
for relativistic electrons the dominant interaction is transverse and not
subject to Debye screening \citep{HP93}. Second, the estimate assumes
a thermal (FD) distribution of electrons, which suppresses collisions except
near the top of the Fermi sea; however, the distribution is nonthermal
and the thermal assumption is inappropriate. Third, the quantization of the perpendicular 
motion is not taken into account. The first two effects tend to underestimate the
actual collision rate, but it seems unlikely that the underestimate is
large enough for electron-electron collisions to dominate over
electron-quark collisions. A more quantitative treatment of each
of these effects is needed in a more detailed discussion.

Collisions with quarks below the electron layer tend to thermalize
the electrons, in particular, forming a high-energy thermal tail that
extends to the escape energy. We develop a simple model for
the production of escaping electrons due to collisions with quarks;
the model involves a non-zero flux in energy space feeding electrons
into the energy range where they can escape.
We interpret the resulting escape in terms of thermionic-like emission.
The rate of pair production is given by (\ref{Richardson}), and
is very sensitive to the work function, $W$, which is identified as the
difference between the escape energy and the maximum energy,
$eV$, of the electrons produced by SPC. Although $W$ is poorly
determined by models for SSs, it seems likely that the rate
(\ref{Richardson}) is well below the BB emission rate (which it
cannot exceed), supporting the suggestion that SSs may be better described as 
`silver spheres' than as black bodies  \citep{AFO86}.

The direct effect of the magnetic field on the pair emission is relatively minor.
It affects the rate of SPC in vacuo, but the rate is suppressed by degeneracy
and regulated by the collisions such that the net rate of SPC is unrelated to
the intrinsic rate in vacuo. The most important consequence of including the
magnetic field is the quantization of the electron states 
into discrete Landau levels. Although a formal theory exists for 
electron-electron collisions in the magnetized case \citep{L81,SM87},
semi-quantitative estimates that would be useful in the present
context are not available. 

We conclude that treating pair emission from unmagnetized SSs
assuming a FD distribution of electrons underestimates
the suppression of SPC due to degeneracy, and overestimates the
luminosity in pairs. The suppression of SPC is self-regulatory, and it adjusts
so that SPC occurs at just the rate needed to replace electrons that
escape. The simplest useful model for pair emission from SSs involves
treating it as thermionic emission, but a quantitative estimate requires
an estimate of the work function, which is poorly determined.

\section*{Acknowledgments} 
We thank Vladimir Usov for suggesting this problem and for helpful correspondence,
and an anonymous referee for perceptive, constructive criticisms.


\appendix
\section{Pair creation in an electromagnetic wrench}

The Lorentz 
invariants associated with a static electromagnetic field are 
$\mathbf{E}\cdot \mathbf{B}$ and $|\mathbf{E}|^2 - |\mathbf{B}|^2$. 
An electromagnetic field can be an electromagnetic wrench 
($\mathbf{E}\cdot \mathbf{B} \neq 0$), an electrostatic field 
($\mathbf{E}\cdot \mathbf{B}=0, |\mathbf{E}|>|\mathbf{B}|$) or a 
magnetostatic field ($\mathbf{E}\cdot \mathbf{B}=0, 
|\mathbf{B}|>|\mathbf{E}|$). For a magnetostatic field, there exist 
frames in which only a magnetic field is present, and such a field is 
strictly stable against SPC. For an electrostatic 
field there exist  frames in which there is no magnetic field. An 
electrostatic field is intrinsically unstable to decay due to
SPC \citep{S51}.  A similar effect occurs for an 
electromagnetic wrench, for which there exists a frame where 
$\mathbf{E}\parallel \mathbf{B}$. In this frame the electric field 
decays, creating pairs at the rate per unit volume per unit time 
\citep{DL76}
\be
w_{\parallel} = \frac{e^2|EB|}{4\pi^2\hbar^2} \sum_{k=1}^{\infty} 
\frac{1}{k} \coth \left[ k \pi \frac{|cB|}{|E|} \right]
\exp
\left[-k \pi\frac{E_c}{|E|} \right],
\label{Edecay}
\ee
with $E_c = m^2c^3/e\hbar$. 

\section{Escaping flux of electrons}

Writing $\tau=\nu_ct$, ${\tilde p}=pc/kT$, the effect of collisions between quarks and electrons
in the nondegenerate tail of the electron distribution at $pc>\ve_\F$ in the ground state, can be approximated by
\be
\frac{\partial{\tilde f}({\tilde p})}{\partial\tau}=\frac{\partial}{\partial{\tilde p}}
\left\{d({\tilde p})\left[\frac{\partial {\tilde f}({\tilde p})}{\partial {\tilde p}}+{\tilde f}({\tilde p})\right]\right\},
\label{collisions}
\ee
where $\tilde f$ is the distribution function, and
$d({\tilde p})$ is determined by the $p$-dependence of the collisions. The two
independent stationary solutions of (\ref{collisions}) are
\be
{\tilde f}_1({\tilde p})=e^{-{\tilde p}},
\qquad
{\tilde f}_2({\tilde p})=e^{-{\tilde p}}\int^{{\tilde p}}\frac{d{\tilde p}'\,e^{{\tilde p}'}}{d({\tilde p}')},
\label{f1f2}
\ee
and a general solution is ${\tilde f}({\tilde p})=A_1{\tilde f}_1({\tilde p})+A_2{\tilde f}_2({\tilde p})$, where $A_{1,2}$ are constants. The solution ${\tilde f}_1({\tilde p})$ contains no flux in momentum space, and the solution ${\tilde f}_2({\tilde p})$ contains a constant flux, from ${\tilde p}=0$ to an unspecified upper momentum. One can include escape of electrons at $p\ge p_0$, or ${\tilde p}\ge{\tilde p}_0$, by requiring ${\tilde f}({\tilde p}_0)=0$. The rate electrons escape is then proportional to the flux in momentum space, determined by
\be
-\int^{{\tilde p}_0}d{\tilde p}\frac{\partial {\tilde f}({\tilde p})}{\partial\tau}=-A_2.
\label{esc}
\ee
(The flux is from ${\tilde p}=0$ to ${\tilde p}={\tilde p}_0$.) The constants $A_{1,2}$ are determined by 
${\tilde f}({\tilde p}_0)=0$ and say ${\tilde f}({\tilde p}_\F)=1$, where $p_\F$ is the Fermi momentum. This gives
\be
A_2=-\left[e^{-{\tilde p}_\F}\int_{{\tilde p}_\F}^{{\tilde p}_0}
\frac{d{\tilde p}\,e^{{\tilde p}}}{d({\tilde p})}
\right]^{-1}.
\label{A2}
\ee
In the special case in which $d({\tilde p})$ does not depend on momentum, $d({\tilde p})\to1$, corresponding to reflection off hard spheres, (\ref{A2}) gives $A_2\approx e^{-W/kT}$, where $W=(p_0-p_\F)c$ is assumed much larger than $kT$. 

\end{document}